\documentclass[a4paper,10pt,oncolumn,twoside]{cpc-hepnp}
\usepackage{CJK,upgreek,fancyhdr}
\usepackage{multicol}
\usepackage{multirow}
\usepackage{graphicx}
\usepackage{booktabs}
\usepackage{amssymb,bm,mathrsfs,bbm,amscd}
\usepackage[tbtags]{amsmath}
\usepackage{lastpage}
\usepackage[english]{babel}
\usepackage{epsfig}
\usepackage{graphicx}
\usepackage{epstopdf}
\usepackage{graphbox}

\begin{document}

\fancyhead[c]{\small Chinese Physics C~~~Vol. xx, No. x (2023) xxxxxx}
\fancyfoot[C]{\small 010201-\thepage}
\footnotetext[0]{Received \today}

\title{Consistency of Pantheon+ supernovae with a large-scale isotropic universe\thanks{Supported by the National Natural Science Fund of China (grant nos. 12147102 and 12275034), and the Fundamental Research Funds for the Central Universities of China (grant no. 2023CDJXY-048).}}

\author{Li Tang$^{1,4}$
\quad Hai-Nan Lin$^{2,3;1)}$\email{linhn@cqu.edu.cn}
\quad Liang Liu$^{1,4}$
\quad Xin Li$^{2,3}$}

\maketitle
\hspace{1cm}

\address{$^1$ Department of Math and Physics, Mianyang Teachers' College, Mianyang 621000, China\\
$^2$ Department of Physics, Chongqing University, Chongqing 401331, China\\
$^3$ Chongqing Key Laboratory for Strongly Coupled Physics, Chongqing University, Chongqing 401331, China\\
$^4$ Research Center of Computational Physics, Mianyang Teachers' College, Mianyang 621000, China}

\begin{abstract}
  We investigate the possible anisotropy of the universe using the most up-to-date type Ia supernovae, i.e. the Pantheon+ compilation. We fit the full Pantheon+ data with the dipole-modulated $\Lambda$CDM model, and find that it is well consistent with a null dipole. We further divide the full sample into several subsamples with different high-redshift cutoff
  $z_c$. It is shown that the dipole appears at $2\sigma$ confidence level only if $z_c\leq 0.1$, and in this redshift region the dipole is very stable, almost independent of the specific value of $z_c$. For $z_c=0.1$, the dipole amplitude is $D=1.0_{-0.4}^{+0.4}\times 10^{-3}$, pointing towards $(l,b)=(334.5_{\ -21.6^{\circ}}^{\circ +25.7^{\circ}},16.0_{\ -16.8^{\circ}}^{\circ +27.1^{\circ}})$, which is about $65^{\circ}$ away from the CMB dipole. This implies that the full Pantheon+ is consistent with a large-scale isotropic universe, but the low-redshift anisotropy couldn't be purely explained by the peculiar motion of the local universe.
\end{abstract}

\begin{keyword}
type Ia supernovae \---  anisotropy  \---  cosmological parameters
\end{keyword}


\footnotetext[0]{\hspace*{-3mm}\raisebox{0.3ex}{$\scriptstyle\copyright$}2019
Chinese Physical Society and the Institute of High Energy Physics
of the Chinese Academy of Sciences and the Institute
of Modern Physics of the Chinese Academy of Sciences and IOP Publishing Ltd}%

\section{Introduction}\label{sec:introduction}

The standard cosmological model, known as the $\Lambda$CDM model, is based on the theory of general relativity and the assumption of cosmological principle. Thereinto, the cosmological principle extends the Copernican principle and postulates that the universe is statistically homogeneous and isotropic on large scales. This assumption on the large-scale structure of the universe is supported by the approximate isotropy of cosmic microwave background (CMB) radiation observed by e.g. the Wilkinson Microwave Anisotropy Probe (WMAP) \cite{WMAP:2012fli,WMAP:2012nax} and the Planck satellites \cite{Planck:2013pxb,Planck:2015fie}. However, the cosmological principle also confronts severe challenges from other observations on the large-scale structure. These challenges include the alignment of quasar polarization vectors on large scales \cite{Hutsemekers:2005iz}, the spatial variations of fine-structure constant \cite{King:2012id,Mariano:2012wx}, and the alignments of low multipoles in CMB angular power spectrum \cite{Lineweaver:1996xa,Tegmark:2003ve,Frommert:2009qw}, and so on. These abnormal phenomena suggest that our universe may not be isotropic but instead possess anisotropic characteristics. In addition, the largest observed anisotropy in the CMB is the dipole with an amplitude of approximately $10^{-3}$, pointing towards the direction $(l,b)=(264^{\circ}, 48^{\circ})$ in the galactic coordinates \cite{Lineweaver:1996xa,Planck:2013kqc,Saha:2021bay}. All these imply that our universe may deviate from the standard $\Lambda$CDM model. Particularly, the Hubble tension problem \cite{Riess:2018uxu,Riess:2021jrx}, i.e. the discrepancy between the Hubble constant values measured from the local distance ladders and from the CMB, further poses severe challenges to the standard cosmological model.

Since the discovery of a dipole signal in a dataset of 100 type Ia supernovae (SNe Ia), which is consistent with the CMB dipole at a confidence level of more than 2$\sigma$ \cite{Bonvin:2006en}, various samples of SNe Ia have been used to investigate the potential anisotropy of the universe \cite{Antoniou:2010gw,Cai:2013lja,Zhao:2013yaa,Chang:2014nca,Chang:2014wpa,Wang:2014vqa,Lin:2016jqp,Lin:2015rza,Wang:2017ezt,Andrade:2017iam,Chang:2017bbi,Deng:2018yhb,Colin:2019opb,Chang:2019utc,Horstmann:2021jjg}. Using the Union2 dataset, Antoniou $\&$ Perivolaropoulos \cite{Antoniou:2010gw} employed the hemisphere comparison method and identified a maximum dark energy dipole direction pointing towards $(l,b)=(309^{\circ},18^{\circ})$. Utilizing the same dataset but employing the dipole fitting method, Mariano $\&$ Perivolaropoulos \cite{Mariano:2012wx} found a dark energy dipole direction pointing towards $(l, b) = (309.4^{\circ}, -15.1^{\circ})$. Furthermore, Zhao et al. \cite{Zhao:2013yaa} divided the Union2 dataset into 12 subsamples and discovered a dipole in the deceleration parameter at a confidence level exceeding 2$\sigma$. A dipole at the level of $2\sigma$ was also found in the Union2.1 dataset \cite{Wang:2014vqa,Lin:2016jqp}. With the JLA sample, Lin et al. \cite{Lin:2015rza} however found no significant deviations from the cosmological principle, see also \cite{Wang:2017ezt,Chang:2017bbi,Andrade:2017iam}. The much larger Pantheon compilation was also utilized to probe the dipole of the universe. Chang et al. \cite{Chang:2019utc} found that the dipole anisotropy is very weak in three dipole-modulated cosmological models. On the other hand, Horstmann et al. \cite{Horstmann:2021jjg} found that the direction of solar motion inferred from Pantheon is consistent with the dipole direction inferred from CMB. In conclusion, whether our universe is anisotropic or not is still in extensive debate. Therefore, it is important to conduct more thorough investigations using high-quality SNe Ia samples before drawing firm conclusions.

Recently, Scolnic et al. \cite{Scolnic:2021amr} compiled the most up-to-date dataset of SNe Ia, which includes 1701 light curves from 1550 unique SNe Ia observed in 18 different surveys. The dataset covers a wide redshift range of $0.001<z<2.3$. Particularly, the SNe Ia at $z<0.01$ in Pantheon+ sample is valuable for analyzing anisotropy in the low-redshift universe. Sorrenti et al. \cite{Sorrenti:2022zat} investigated the dipole signal using subsamples of the full Pantheon+, based on different low-redshift cutoffs. Their results show that the amplitudes of the dipole signal roughly agree with the CMB dipole, independent of the redshift cutoffs. However, the dipole directions significant different from the CMB dipole. In another study, McConville $\&$ Colg$\acute{a}$in \cite{McConville:2023xav} analyzed the anisotropic variation of the Hubble constant using two subsamples of the full Pantheon+ in the redshift ranges $0.0233 < z < 0.15$ and $0.01 < z < 0.7$, respectively. They found that the Hubble constant is significantly larger in a hemisphere encompassing the CMB dipole direction, but the variation of the Hubble constant is not large enough to reconcile the Hubble tension problem.

These recent investigations highlight the importance of exploring anisotropy using the most updated and comprehensive SNe Ia datasets, such as the Pantheon+ sample, and reveal interesting discrepancies in the dipole directions when compared to the CMB observations. Further analysis and scrutiny are necessary to fully understand the implications and to establish more robust conclusions regarding the presence of anisotropy in the universe. In this paper, we will further investigate the possible anisotropy hiding in the Pantheon+ compilation with the dipole fitting method. Different from Sorrenti et al. \cite{Sorrenti:2022zat}, who attributes the anisotropy to the peculiar motion of our solar system with respect to CMB, we directly parameterize the anisotropy as the dipole form. If the anisotropy purely originates from the peculiar motion, then the dipole of Pantheon+ is expected to be aligned with the dipole of CMB, and the dipole signal should be more obvious at low redshift. Therefore, we also investigate the anisotropy by dividing the full sample into several subsamples according to redshift. It should be noted that, our method to divide the subsamples is different from Sorrenti et al. \cite{Sorrenti:2022zat}. We divide the full sample into subsamples using various high-redshift cutoffs, in contrast to the low-redshift cutoffs used in Sorrenti et al. \cite{Sorrenti:2022zat}. Comparing with the high-redshift SNe, the low-redshift SNe are expected to be more substantially affected by the peculiar motion. Therefore, our research places greater emphasis on the low-redshift subset.

The rest parts of this paper are arranged as follows: In Section 2, we introduce the data and methodology that are involved to test the anisotropy of the universe. The results are illustrated in Section 3. Finally, discussion and conclusions are given in Section 4.

\section{Data and methodology}\label{sec:data}

\begin{figure*}[htbp]
 \centering
 \includegraphics[width=0.8\textwidth]{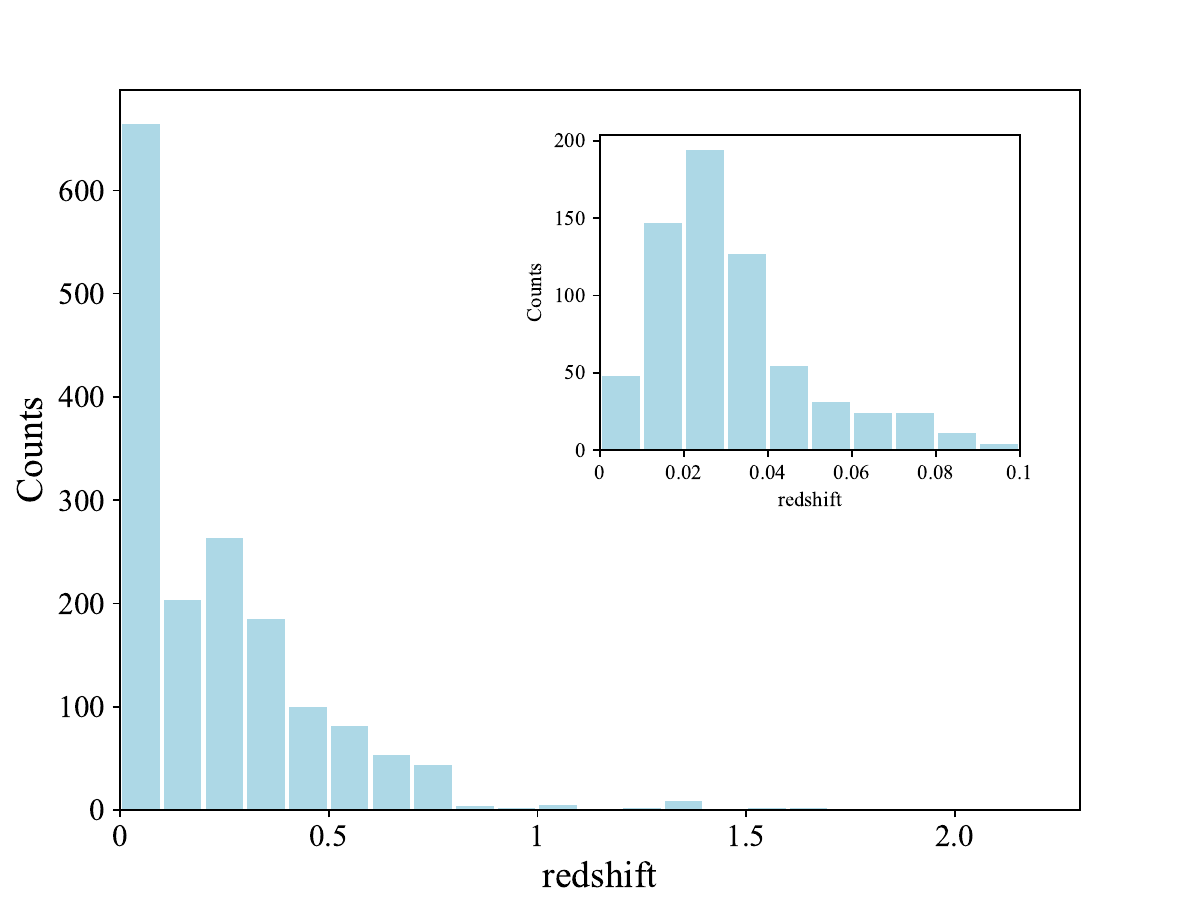}
 \caption{The redshift distribution of the Pantheon+ SNe. Data points are very sparse at $z>0.8$. The inset is the redshift distribution of the low-redshift ($z<0.1$) SNe.}\label{fig:redshift}
\end{figure*}

\begin{figure*}[htbp]
 \centering
 \includegraphics[width=0.8\textwidth]{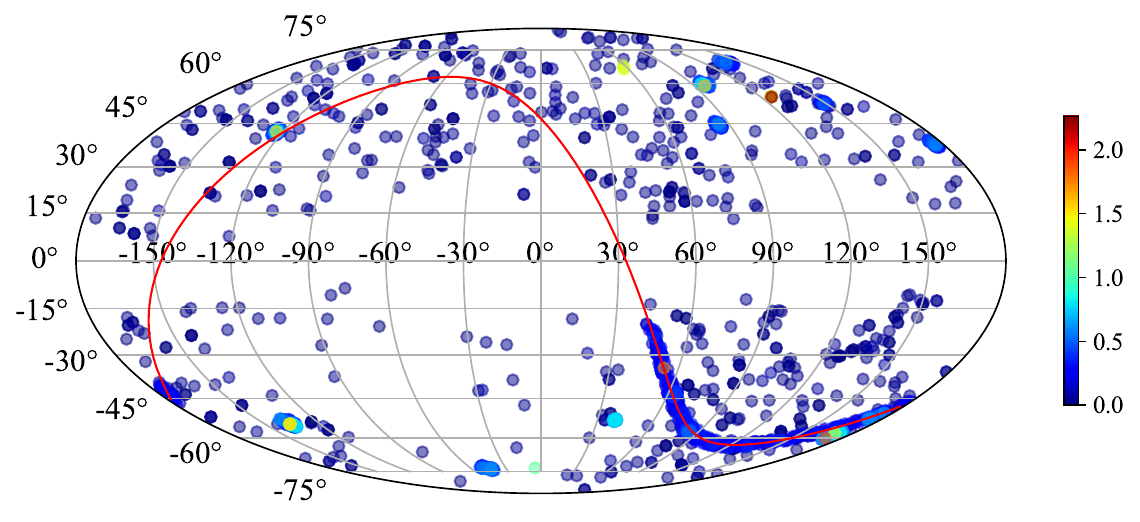}
 \caption{The sky position of the Pantheon+ SNe in the galactic coordinates, colored according to the redshift. The red-solid line is the celestial equator.}\label{fig:sky}
\end{figure*}

SNe Ia serve as ideal standard candles and are widely utilized to constrain the cosmological parameters, especially to investigate the anisotropy of the universe. Recently, Scolnic et al. \cite{Scolnic:2021amr} published the most up-to-date compilation of SNe Ia data, known as the Pantheon+ compilation, which is the updated version of the previous Pantheon compilation \cite{Pan-STARRS1:2017jku}. One major difference between the Pantheon+ and the Pantheon is that the former contains much more low-redshift SNe than the latter, thus allows us to probe the low-redshift universe more thoroughly. The Pantheon+ compilation consists of 1701 light curves from 1550 unique SNe Ia, covering a redshift range of $0.001<z<2.3$. The redshift distribution of the Pantheon+ dataset is illustrated in Figure \ref{fig:redshift}. This figure demonstrates that the majority of SNe are concentrated at low redshift range ($z<0.1$), while at high redshift range ($z>0.8$) the data points are very sparse. To provide a clearer view of this concentration, the inset of Figure \ref{fig:redshift} specifically highlights the redshift distribution below 0.1. Additionally, in Figure \ref{fig:sky} we plot the sky positions of the Pantheon+ SNe in the galactic coordinates, which reveals an inhomogeneous distribution, with a large number of data points concentrate near the celestial equator (see the red-solid line in Figure \ref{fig:sky}).

The observed distance modulus $\mu_{\rm obs}$ is derived from the light curve parameters using a modified version of the Tripp formula \cite{Tripp:1997wt}:
\begin{equation}\label{eq:mu_obs}
\mu_{\rm obs}=m_B-M+\alpha x_1-\beta c_1-\delta_{\rm bias}+\delta_{\rm host},
\end{equation}
where $m_B$ represents the apparent magnitude in the B-band, and $M$ corresponds to the absolute magnitude. The parameter $x_1$ is associated with the stretch of the light curve width, and $c_1$ represents the color parameter, which is influenced by both intrinsic color and dust effects. The nuisance parameters $\alpha$ and $\beta$ are used to account for the link between the stretch $x_1$ and color $c_1$ with the luminosity. The terms $\delta_{\rm bias}$ and $\delta_{\rm host}$ represent correction terms. Thereinto, $\delta_{\rm bias}$ accounts for the selection biases from simulations, $\delta_{\rm host}$ considers the contribution of the host galaxy mass of the SNe Ia. After calibration using the BEAMS with Bias Corrections (BBC) method \cite{Kessler:2016uwi}, the Pantheon+ dataset provides the corrected magnitude $m_{\rm obs}$, along with the corresponding covariance matrix. The observed distance modulus is then expressed as $\mu_{\rm obs}=m_{\rm obs}-M$. The details of the Pantheon+ dataset can be found in Scolnic et al. \cite{Scolnic:2021amr}.

In a spatially flat universe, the dimensionless distance modulus at a given redshift $z$ can be expressed as
\begin{equation}\label{eq:mu_iso}
\mu(z)=5\log_{10}\frac{d_L(z)}{\rm Mpc}+25,
\end{equation}
where $d_L$ represents the luminosity distance and is measured in units of Mpc. In the framework of the standard $\Lambda$CDM model, the luminosity distance takes the following form
\begin{equation}\label{eq:D_L}
d_L(z)=(1+z)\frac{c}{H_0}\int^z_0\frac{dz}{\sqrt{\Omega_{\rm M}(1+z)^3+\Omega_{\Lambda}}},
\end{equation}
where $c$ denotes the speed of light, $H_0$ represents the Hubble constant, which is typically parameterized in terms of a dimensionless parameter $h\equiv H_0/(100~{\rm km/s/Mpc})$, $\Omega_{\rm M}$ and $\Omega_{\Lambda}=1-\Omega_{\rm M}$ denote the dimensionless densities of matter and dark energy, respectively.

To investigate the anisotropy of the universe, the dipole fitting method, which was first introduced by Mariano et al. \cite{Mariano:2012wx} is utilized. This method fits the observational data using a dipole model directly. According to the dipole fitting method, a dipole modulation is introduced to the theoretical distance modulus in the isotropic $\Lambda$CDM model, given by the following form,
\begin{equation}\label{eq:mu_D}
\mu_D(z)=\mu_{\rm iso}(z)\left[1+D(\bm{\hat{n}} \cdot \bm{\hat{p}})\right].
\end{equation}
Here, $\mu_{\rm iso}(z)$ represents the distance modulus in the isotropic $\Lambda$CDM model determined by equation (\ref{eq:mu_iso}), $D$ is the amplitude of the dipole, $\bm{\hat{n}}$ is the direction of the dipole, and $\bm{\hat{p}}$ is the unit vector pointing towards the SNe Ia. In the galactic coordinates, the two unit vectors $\bm{\hat{n}}$ and $\bm{\hat{p}}$ can be parameterized using the galactic longitude $l$ and latitude $b$ in the following manner
\begin{equation}
\bm{\hat{n}}=\cos(b_0)\cos(l_0)\bm{\hat{i}}+\cos(b_0)\sin(l_0)\bm{\hat{j}}+\sin(b_0)\bm{\hat{k}},
\end{equation}
\begin{equation}
\bm{\hat{p_i}}=\cos(b_i)\cos(l_i)\bm{\hat{i}}+\cos(b_i)\sin(l_i)\bm{\hat{j}}+\sin(b_i)\bm{\hat{k}},
\end{equation}
where $\bm{\hat{i}}$, $\bm{\hat{j}}$, and $\bm{\hat{k}}$ represent the unit vectors along the three axes of Cartesian coordinates, ($l_0$, $b_0$) and ($l_i$, $b_i$) represent the direction of dipole and the direction of the $i$-th SN in the galactic coordinates, respectively. In the Pantheon+ dataset, the SNe positions are provided in the equatorial coordinates. In order to directly compare with other works, we transform the equatorial coordinates $({\rm RA, DEC})$ to the galactic coordinates ($l,b$) using the formulae given in Ref. \cite{duffett-smith_1989}.

The free parameters in our analysis consist of the matter density $\Omega_{\rm m}$, the absolute magnitude $M$, the dimensionless Hubble constant $h$, the dipole amplitude $D$, and the dipole direction $(l_0,b_0)$. Thereinto, the absolute magnitude $M$ and the dimensionless Hubble constant $h$ cannot be simultaneously constrained using SNIa data alone due to their degeneracy. Fortunately, the absolute magnitude $M$ also can be refined through the establishment of an absolute distance scale, employing primary distance anchors such as Cepheids. The Pantheon+ dataset has extended the lower redshift boundary of SNe Ia to 0.001, which encompasses primary distance indicators found in Cepheid host galaxies. The degeneracy between $M$ and $h$ is eliminated by combining the measurements from the SH0ES Cepheid and SNe data. As a result, $M$ and $h$ can be constrained simultaneously. Following the methodology outlined in Brout et al. \cite{Brout2022}, the best-fitting parameters are determined by maximizing the likelihood function, which is related to the $\chi^2$ by
\begin{equation}
-2\ln(\mathcal{L})=\chi^2=\Delta\bm{\mu}^{T}\bm{C}^{-1}\Delta\bm{\mu},
\end{equation}
in which $\bm{C}$ represents the total covariance matrix, which combines both the statistical and systematic covariance matrices. $\Delta\bm{\mu}$ denotes the residual vector of the distance modulus, where the $i$-th element is defined as
\begin{eqnarray}
\Delta\mu_i=
\begin{cases}
\mu_{\rm{obs},\textit{i}}-\mu_{\rm{ceph},\textit{i}} \quad & i\in \rm{Cepheid} \ \rm{hosts},\\
\mu_{\rm{obs},\textit{i}}-\mu_{D,i} \quad & \rm{otherwise}.
\end{cases}
\end{eqnarray}
Here, $\mu_{{\rm{ceph}},i}$ corresponds to the calibrated distance to the host galaxy determined from the Cepheid measurements, while $\mu_{{\rm{obs}},i}$ and $\mu_{D,i}$ are determined by equation (\ref{eq:mu_obs}) and equation (\ref{eq:mu_D}) respectively.

\section{The results}\label{sec:results}

We employ the Markov Chain Monte Carlo (MCMC) method, specifically the affine-invariant MCMC ensemble sampler provided by the publicly available python package \textsf{emcee}\footnote{https://emcee.readthedocs.io/en/stable/} \cite{Foreman-Mackey:2012any}, to perform the parameter fitting. The posterior probability density functions (PDFs) of the free parameters are calculated using this approach. For each parameter, we assume a flat prior distribution as follows: $\Omega_{\rm M}\sim[0,1]$, $M\sim[-21.0,-18.0]$, $h\sim[0.6,0.8]$, $D\sim[0,0.01]$, $l_0\sim[-180^{\circ},180^{\circ}]$\footnote{In convention, $l_0$ is usually taken to be in the range $[0^{\circ},360^{\circ}]$. However, we find that the best-fitting value is near the boundary of this range. So we use the prior $l_0\sim[-180^{\circ}, 180^{\circ}]$ in the MCMC calculation. The results can be converted in to the range of $[0^{\circ},360^{\circ}]$ by adding $360^{\circ}$ to the negative $l_0$, while keeping the positive $l_0$ unchanged.}, and $b_0\sim[-90^{\circ},90^{\circ}]$.

\begin{figure*}[htbp]
 \centering
 \includegraphics[width=0.49\textwidth]{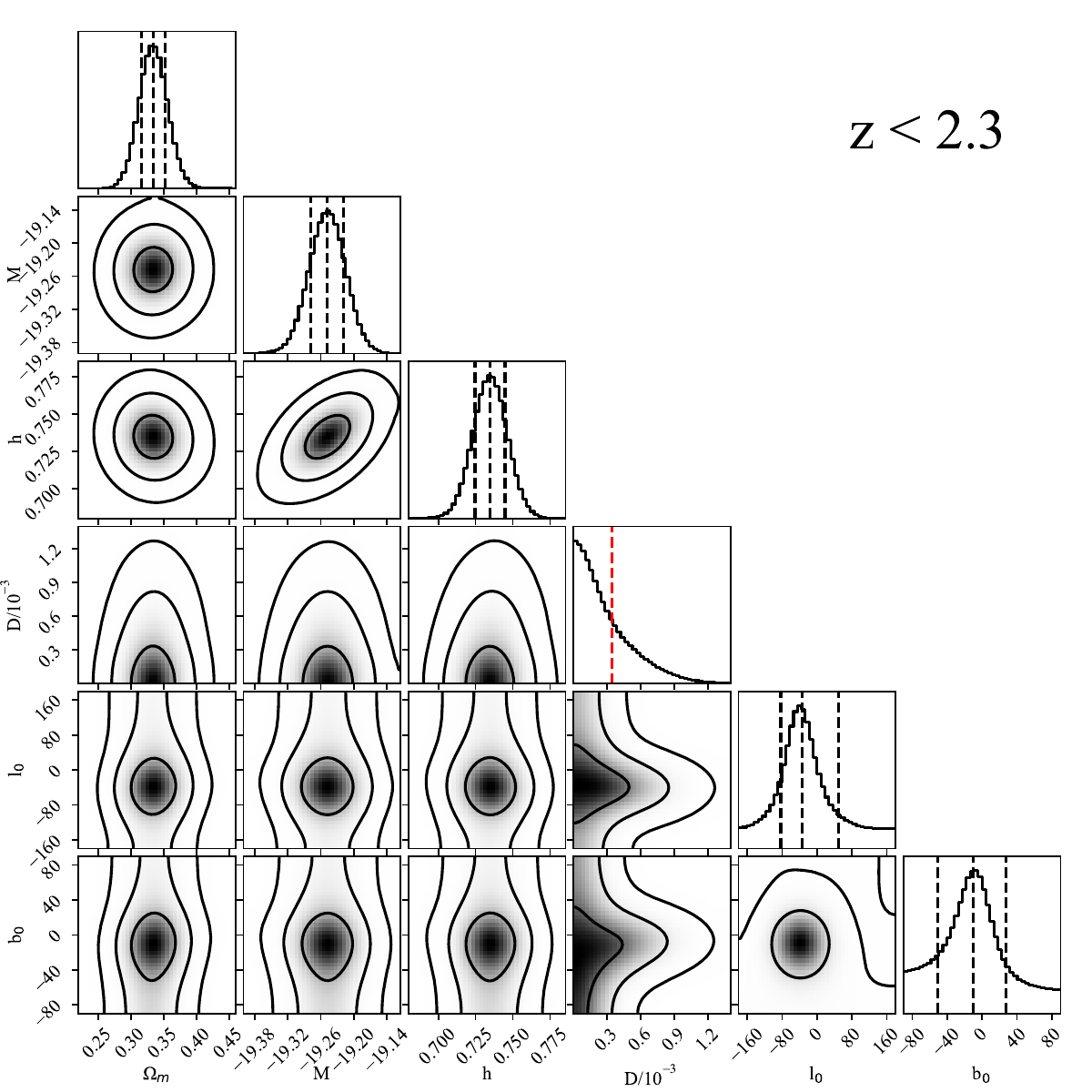}
 \includegraphics[width=0.49\textwidth]{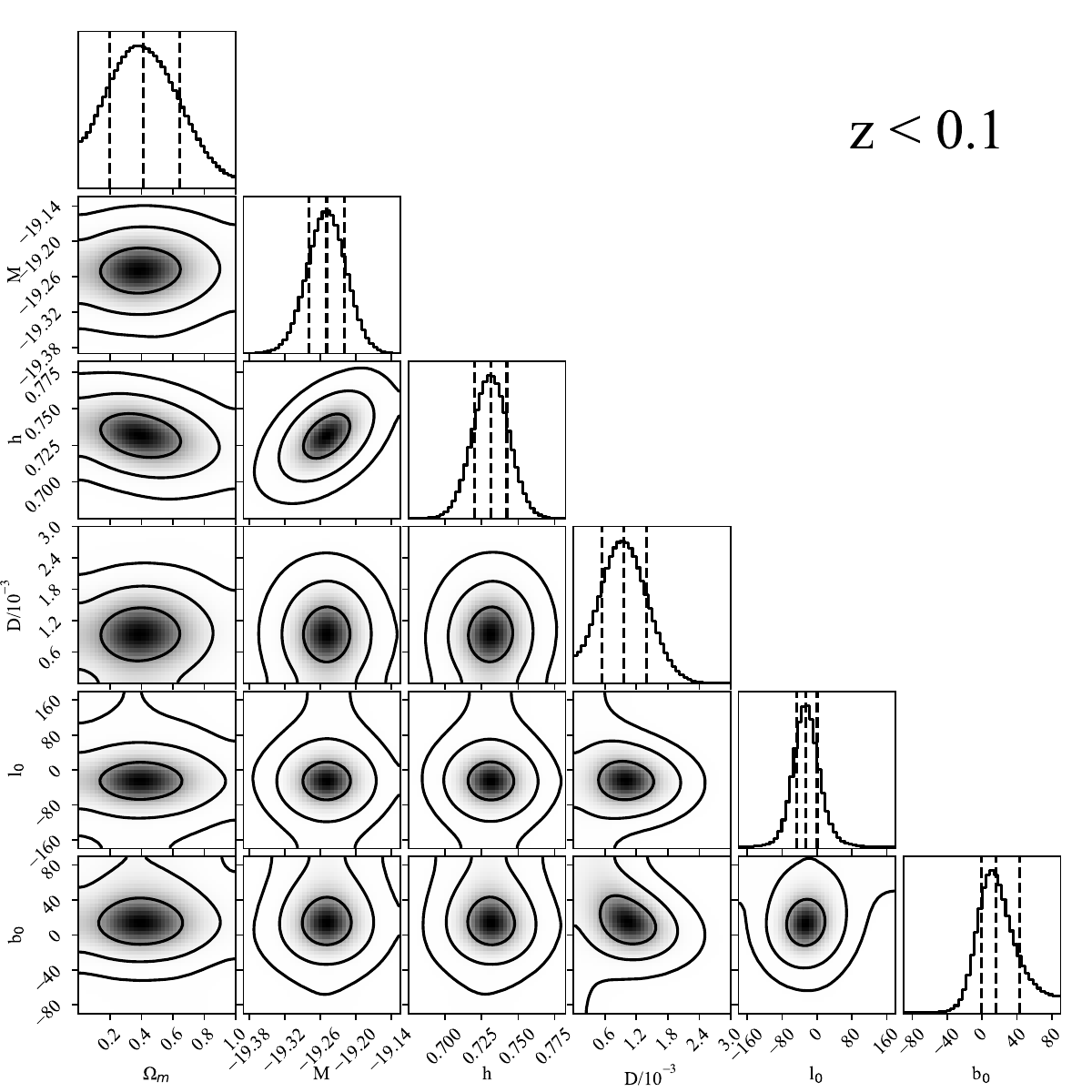}
 \caption{The posterior PDFs of parameters and the 2-dimensional confidence contours constrained from the full Pantheon+ sample (left panel) and the $z<0.1$ subsample (right panel). The black dashed lines represent the median value and $1\sigma$ uncertainty, and the red dashed line is the $1\sigma$ upper limit.}\label{fig:triangle}
\end{figure*}

First, we use the full Pantheon+ sample to constrain the free parameters of the dipole-modulated $\Lambda$CDM model. The full sample contains 1701 data points in the redshift range $z<2.3$, including 77 SNe in galaxies hosting Cepheids. The constraints on the parameters are summarized in the first row of Table \ref{tab:parameters}. The left panel of Figure \ref{fig:triangle} shows the corresponding 2-dimensional confidence contours and 1-dimensional posterior PDFs for the parameters. In the dipole-modulated $\Lambda$CDM model, the background parameters $\Omega_{\rm M}$, $M$, and $h$ are tightly constrained as $\Omega_{\rm M}=0.334^{+0.018}_{-0.018}$, $M=-19.248_{-0.030}^{+0.029}$, and $h=0.735_{-0.010}^{+0.010}$, which are consistent with those obtained from the isotropic flat $\Lambda$CDM model \cite{Brout2022} within 1$\sigma$ uncertainty. Regarding to the dipole component, the anisotropic signal is weak in the full Pantheon+ sample. The 68$\%$ upper limit of the dipole amplitude is constrained to be $D<0.3\times 10^{-3}$, and the dipole direction points towards $(l_0,b_0)=(326.3_{\ -49.5^{\circ}}^{\circ+82.4^{\circ}},-10.4_{\ -40.0^{\circ}}^{\circ+37.9^{\circ}})$. The small upper limit of the dipole amplitude, as well as the large uncertainty on the dipole direction indicate that the full Pantheon+ dataset is well consistent with a large-scale isotropic universe.

\begin{table}[htbp]
\centering
\caption{\small{The best-fitting parameters of the dipole-modulated $\Lambda$CDM model. The uncertainties are given at $1\sigma$ confidence level. For the dipole amplitude, the $68\%$ upper limit is reported when the posterior DPF has no obvious peak. The second column is the number of data points. The galactic longitude $l_0$ is converted into the range of $[0^{\circ},360^{\circ}]$.}}\label{tab:parameters}
\arrayrulewidth=1.0pt
\renewcommand{\arraystretch}{1.3}
{\begin{tabular}{cccccccc} 
\hline\hline 
Sample & $N$ &  $\Omega_m$  & $M$ & $h$     & $D/10^{-3}$   &$l_0[^{\circ}]$  &$b_0[^{\circ}]$\\\hline
$z<2.3$ &1701& $0.334_{-0.018}^{+0.018}$ & $-19.248_{-0.030}^{+0.029}$  & $0.735_{-0.010}^{+0.010}$   & $<0.3$   &$326.3_{-49.5}^{+82.4}$ &$-10.4_{-40.0}^{+ 37.9}$\\
$z<0.7$ &1626& $0.348_{-0.021}^{+0.022}$ & $-19.245_{-0.030}^{+0.030}$  & $0.735_{-0.010}^{+0.010}$   & $<0.4$   &$320.8_{-42.2}^{+64.5}$ &$-12.6_{-35.8}^{+ 29.8}$\\
$z<0.5$ &1492& $0.332_{-0.024}^{+0.025}$ & $-19.244_{-0.030}^{+0.029}$  & $0.736_{-0.010}^{+0.010}$   & $<0.5$   &$320.7_{-40.1}^{+67.3}$ &$-13.1_{-33.5}^{+ 29.4}$\\
$z<0.4$ &1392& $0.341_{-0.029}^{+0.030}$ & $-19.245_{-0.029}^{+0.030}$  & $0.735_{-0.010}^{+0.010}$   & $<0.4$   &$317.2_{-50.5}^{+80.2}$ &$-3.3_{-35.7}^{+ 43.5}$\\
$z<0.3$ &1207& $0.404_{-0.043}^{+0.045}$ & $-19.246_{-0.030}^{+0.029}$  & $0.733_{-0.010}^{+0.010}$   & $<0.7$   &$315.6_{-31.0}^{+46.3}$ &$12.6_{-22.3}^{+ 39.3}$\\
$z<0.2$ &944& $0.446_{-0.073}^{+0.075}$ & $-19.246_{-0.030}^{+0.030}$  & $0.732_{-0.010}^{+0.010}$   & $0.7_{-0.4}^{+0.4}$   &$323.7_{-24.9}^{+35.0}$ &$13.2_{-19.2}^{+33.9}$\\
$z<0.1$ &741& $0.411_{-0.213}^{+0.235}$ & $-19.249_{-0.030}^{+0.030}$  & $0.731_{-0.011}^{+0.011}$   & $1.0_{-0.4}^{+0.4}$   &$334.5_{-21.6}^{+25.7}$ &$16.0_{-16.8}^{+27.1}$\\
$z<0.07$ &702& $0.551_{-0.280}^{+0.263}$ & $-19.250_{-0.030}^{+0.030}$  & $0.729_{-0.011}^{+0.011}$   & $0.8_{-0.4}^{+0.4}$   &$330.9_{-29.1}^{+34.0}$ &$26.0_{-22.5}^{+34.1}$\\
$z<0.04$ &593& $0.344_{-0.245}^{+0.349}$ & $-19.250_{-0.030}^{+0.029}$  & $0.730_{-0.011}^{+0.011}$   & $1.0_{-0.4}^{+0.4}$   &$316.4_{-31.3}^{+34.9}$ &$36.4_{-24.1}^{+31.9}$\\
$z<0.03$ &466& $0.635_{-0.366}^{+0.261}$ & $-19.250_{-0.030}^{+0.029}$  & $0.733_{-0.011}^{+0.011}$   & $1.5_{-0.6}^{+0.6}$   &$322.7_{-21.3}^{+23.5}$ &$16.9_{-17.4}^{+26.4}$\\
$z<0.02$ &272& $0.482_{-0.332}^{+0.346}$ & $-19.252_{-0.030}^{+0.029}$  & $0.723_{-0.012}^{+0.012}$   & $1.4_{-0.9}^{+0.9}$   &$282.7_{-39.2}^{+56.6}$ &$20.9_{-30.3}^{+38.6}$\\
\hline
\end{tabular}}
\end{table}

In addition to investigate the dipole using the full Pantheon+ dataset, we also explore the possible redshift-dependence of the dipole by dividing the dataset into several subsamples. These subsamples are obtained by excluding supernovae with redshift higher than a certain cutoff value $z_{\rm c}$. In other words, a subsample consists of the supernovae with redshift $z<z_c$. The cutoff redshift $z_c$ is chosen from 0.1 to the maximum redshift with equal step size $\Delta z=0.1$. Since in some redshift bins the number of data points are very sparse, as is seen from Figure \ref{fig:redshift}, we only consider the subsamples with number difference larger than 100. We finally find six subsamples, with $z_c=0.1,0.2,0.3,0.4,0.5,0.7$. The number of data points in each subsample is listed in the second column of Table \ref{tab:parameters}. It should be noted that all supernovae that in galaxies hosting Cepheids are included in each subsample, even if their redshift exceeds the cutoff value. This is because these supernovae are not only used to determine the absolute magnitude $M$, but also to eliminate the degeneracy between $M$ and $h$.

\begin{figure*}[htbp]
\begin{minipage}[t]{.32\textwidth}
\vspace{0pt}
\centering
{\includegraphics[width=1.\textwidth]{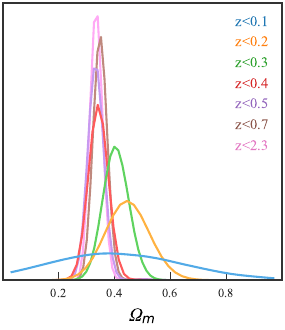}}
\end{minipage}
\begin{minipage}[t]{.32\textwidth}
\vspace{0pt}
\centering
{\includegraphics[width=1.\textwidth]{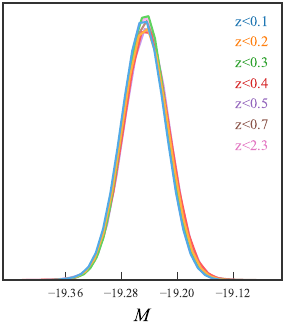}}
\end{minipage}
\begin{minipage}[t]{.32\textwidth}
\vspace{0pt}
\centering
{\includegraphics[width=1.\textwidth]{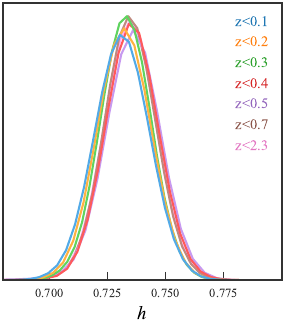}}
\end{minipage}\\
\begin{minipage}[t]{.32\textwidth}
\vspace{0pt}
\centering
{\includegraphics[width=1.\textwidth]{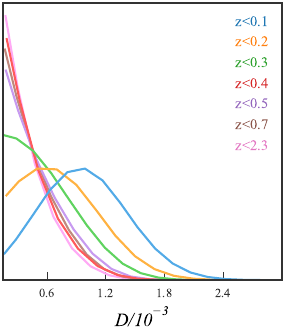}}
\end{minipage}
\begin{minipage}[t]{.32\textwidth}
\vspace{0pt}
\centering
{\includegraphics[width=1.\textwidth]{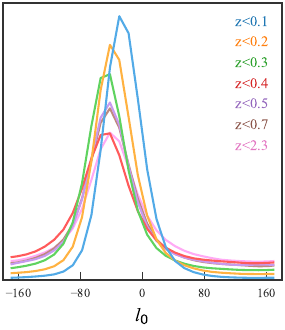}}
\end{minipage}
\begin{minipage}[t]{.32\textwidth}
\vspace{0pt}
\centering
{\includegraphics[width=1.\textwidth]{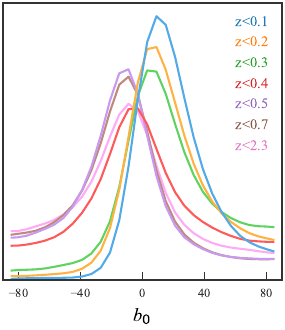}}
\end{minipage}
 \caption{The posterior PDFs of parameters constrained from different subsamples with $z_c\geq 0.1$.}\label{fig:posterior_pdf}
\end{figure*}

We constrain the parameters of the dipole-modulated $\Lambda$CDM model with each subsamples using the same method mentioned above, and the results are summarized in Table \ref{tab:parameters}. Similar to the full Pantheon+ sample, it is found that there is no strong evidence for the presence of dipole anisotropy in the subsamples with a cutoff redshift $z_{\rm c}$ higher than 0.2. However, the dipole signal emerges in the subsamples with $z_{\rm c}\leq0.2$. In the subsample $z<0.2$, the dipole amplitude is constrained as $D=0.7^{+0.4}_{-0.4}\times 10^{-3}$, pointing towards $(l_0,b_0)=(323.7_{\ -24.9^{\circ}}^{\circ+35.0^{\circ}},13.2_{\ -19.2^{\circ}}^{\circ+33.9^{\circ}})$, which deviates from an isotropic universe at $>1\sigma$ confidence level. While in the subsample $z<0.1$, the dipole signal emerges at $>2\sigma$ confidence level, with a dipole amplitude $D=1.0_{-0.4}^{+0.4}\times 10^{-3}$, pointing towerds $(l_0,b_0)=(334.5_{\ -21.6^{\circ}}^{\circ+25.7^{\circ}},16.0_{\ -16.8^{\circ}}^{\circ+27.1^{\circ}})$. The right panel of Figure \ref{fig:triangle} shows the corresponding 2-dimensional confidence contours and 1-dimensional posterior PDFs for the parameters with this subsample. To facilitate comparison, we also plot the posterior PDFs of the parameters constrained from different subsamples in Figure \ref{fig:posterior_pdf}. As can be seen, the significance of dipole signal progressively increases as the decreasing of $z_c$, while the dipole direction is relatively stable.

\begin{figure*}[htbp]
\begin{minipage}[t]{.32\textwidth}
\vspace{0pt}
{\includegraphics[width=1.\textwidth]{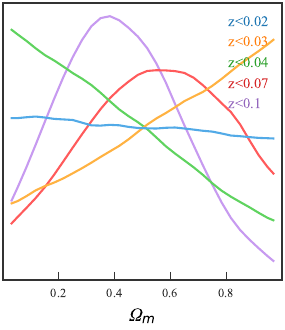}}
\end{minipage}
\begin{minipage}[t]{.32\textwidth}
\vspace{0pt}
{\includegraphics[width=1.\textwidth]{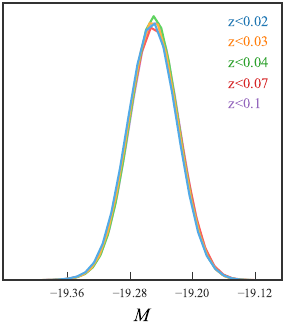}}
\end{minipage}
\begin{minipage}[t]{.32\textwidth}
\vspace{0pt}
{\includegraphics[width=1.\textwidth]{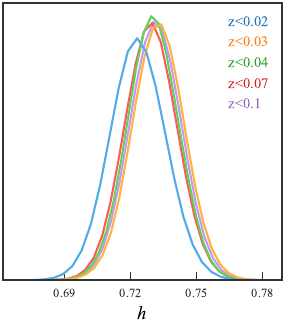}}
\end{minipage}\\
\begin{minipage}[t]{.32\textwidth}
\vspace{0pt}
{\includegraphics[width=1.\textwidth]{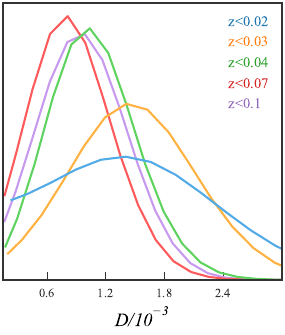}}
\end{minipage}
\begin{minipage}[t]{.32\textwidth}
\vspace{0pt}
{\includegraphics[width=1.\textwidth]{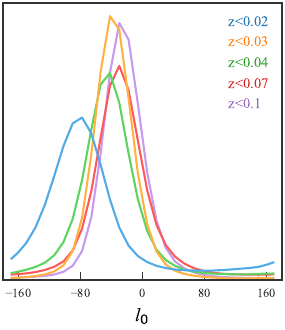}}
\end{minipage}
\begin{minipage}[t]{.32\textwidth}
\vspace{0pt}
{\includegraphics[width=1.\textwidth]{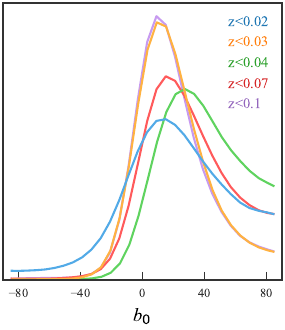}}
\end{minipage}
 \caption{The posterior PDFs of parameters constrained from different subsamples with $z_c\leq 0.1$.}\label{fig:posterior_pdf2}
\end{figure*}

We note that more than one-third of data points have redshift below $0.1$, see Figure \ref{fig:redshift}. In light of the clear dipole signal in the low-redshift range, we perform a thorough examination of the dipole-modulated $\Lambda$CDM model with the low-redshift data points by dividing it into several redshift bins. To achieve this, we further divide the $z<0.1$ data points into several subsamples using the similar method as before, but with a smaller redshift interval of 0.01. We also only consider the subsamples with number difference larger than 100. We finally find four subsamples: $z<0.02$, $z<0.03$, $z<0.04$, $z<0.07$. The constraining results from these subsamples are summarized in the last four columns of Table \ref{tab:parameters}. Because the luminosity distance (see equation (\ref{eq:D_L})) is insensitive to the matter density parameter $\Omega_{\rm M}$ at low-redshift region, this parameter couldn't be tightly constrained. The constraints on the other parameters remain to be consistent across all the subsamples. Especially, the parameters $M$ and $h$ are almost independent of the subsamples. This indicates that the dipole-modulated $\Lambda$CDM model provides stable parameter estimation in the low-redshift range. The posterior PDFs of the parameters, constrained from different low-redshift subsamples, are displayed in Figure \ref{fig:posterior_pdf2}. These plots further support the existence of an anisotropic signal at low-redshift region. Notably, the dipole parameters are stable, remaining nearly independent of the specific value of the cutoff redshift $z_{\rm c}$, although the uncertainty is large for the lowest-redshift subsample ($z<0.02$).

\begin{figure*}[htbp]
 \centering
 \includegraphics[width=0.6\textwidth]{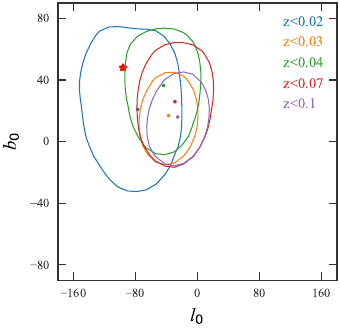}
 \caption{The dipole directions of different subsampes in the sky of galactic coordinates. The contours represent the 1$\sigma$ uncertainty regions of the dipole directions. For comparison, the CMB dipole direction is also shown as the red star.}\label{fig:all_l0_b0}
\end{figure*}

Figure \ref{fig:all_l0_b0} shows the dipole directions constrained from the low-redshift subsamples. In this figure, the contours represent the 1$\sigma$ uncertainty regions of the dipole directions. For comparison, the CMB dipole direction pointing towards $(l,b)=(264^{\circ},48^{\circ})$ \cite{Saha:2021bay} is also shown as the red star. From this figure, we can clearly see that the dipole directions obtained from different subsamples are consistent with each other. It is noteworthy that the dipole directions of all the low-redshift subsamples, except for the lowest-redshift subsample ($z<0.02$), deviate from the CMB dipole at more than $1\sigma$ confidence level. The consistency between the dipole directions of the $z<0.02$ subsample and the CMB is mainly due to the large uncertainty of the former. This implies that the anisotropic signal underlying the low-redshift Pantheon data couldn't be purely explained by the peculiar motion of the local universe.

\section{Discussion and conclusions}\label{sec:conclusions}

In this paper, we have investigated the possible anisotropy of the universe using the most up-to-date SNe Ia dataset, i.e. the Pantheon+ compilation. We phenomenally constructed a dipole-corrected Hubble diagram based on the spatially flat $\Lambda$CDM model, and fitted it to the Pantheon+ compilation. We found that the full Pantheon+ compilation is well consistent with an isotropic universe, with the $1\sigma$ upper limit of dipole amplitude $D<0.3\times 10^{-3}$. To check the possible redshift dependence of the result, we fitted our model with the low-redshift subsamples ($z<z_c$) of Pantheon+, where $z_c$ is the cutoff value. It was shown that the anisotropic signal exists only if $z_c\leq 0.2$. Especially for $z_c=0.1$, the anisotropic signal is at the significance level of $2\sigma$, with the dipole amplitude $D=1.0_{-0.4}^{+0.4}\times 10^{-3}$, and the dipole direction $(l,b)=(334.5_{\ -21.6^{\circ}}^{\circ+25.7^{\circ}}, 16.0_{\ -16.8^{\circ}}^{\circ+27.1^{\circ}})$. This direction is $65^\circ$ away from the CMB dipole, $(l,b)=(264^{\circ},48^{\circ})$ \cite{Lineweaver:1996xa}. If one naively assumes that the anisotropy is induced by the peculiar velocity, one may expect that the dipole of SNe is aligned with the dipole of CMB. Therefore, the dipole of Pantheon+ seen at $z<0.1$ couldn't be purely explained by the peculiar motion of the local universe.

The anisotropy of the universe has been extensively investigated using different groups of SNe Ia. For instance, Mariano $\&$ Perivolaropoulos \cite{Mariano:2012wx} found a dipole in Union2 dataset, with an amplitude $A = (1.3 \pm 0.6) \times 10^{-3}$, pointing towards $(l, b) = (309.4^{\circ} \pm 18.0^{\circ}, -15.1^{\circ} \pm 11.5^{\circ})$. Wang $\&$ Wang \cite{Wang:2014vqa} found a dipole with amplitude $A = (1.37 \pm 0.57) \times 10^{-3}$, pointing towards $(l, b) = (309.2^{\circ} \pm 15.8^{\circ}, -8.6^{\circ} \pm 10.5^{\circ})$ in Union2.1. Using JLA dataset, Lin et al. \cite{Lin:2015rza} found that the dipole is well consistent with null, with the amplitude $A < 1.98 \times 10^{-3}$, pointing towards $(l, b) = (316_{\ -110^{\circ}}^{\circ+107^{\circ}}, -5_{\ -60^{\circ}}^{\circ+41^{\circ}})$. Zhao et al. \cite{Zhao:2019azy} found that there is no evidence for the dipole anisotropy in Pantheon, with amplitude $A < 1.16 \times 10^{-3}$, pointing towards $(l, b) = (306_{\ -125^{\circ}}^{\circ+83^{\circ}}, -34_{\ -55^{\circ}}^{\circ+17^{\circ}})$. In this paper, we also found that the Pantheon+, an updated version of the Pantheon, is well consitent with an isotropic universe, with the dipole amplitude $A < 0.3\times 10^{-3}$, pointing towards $(l,b) = (326.3_{\ -49.5^{\circ}}^{\circ+82.4^{\circ}}, -10.4_{\ -40.0^{\circ}}^{\circ+37.9^{\circ}})$. The dipoles of Union2 and Union2.1 are at the level of $\sim 2\sigma$, while the dipoles of JLA, Pantheon and Pantheon+ are well consistent with null. We note that the redshift range of the former two datasets is $0<z<1.4$, while the redshift range of the later three datasets is $0<z<2.3$. However, the number of SNe with $z>1.4$ is very small. Therefore, the different results are not expected to be caused by the extension of redshift. One reason for the difference may be that the correlation between SNe in Union2 and Union2.1 have been ignored, while the covariance matrix of JLA, Pantheon and Pantheon+ has been full taken into considered. Another reason may be that the number of data points almost tripled from Union2 ($N=557$) to Pantheon+ ($N=1701$). With the enlargement of data sample, the anisotropic signal gradually vanishes.

The dipole anisotropy of the Pantheon+ has also been studied by Sorrenti et al. \cite{Sorrenti:2022zat}. In their paper, the authors attribute the anisotropy to the peculiar motion of our solar system with respect to the CMB frame. The peculiar motion of our solar system induces a redshift correction for each SN, with depends on the position of SN on the sky. They found the peculiar velocity $v_0=328_{-42}^{+35}~{\rm km/s}$, pointing towards $({\rm RA,DEC})=(139.4_{\ -8.0^{\circ}}^{\circ+7.2^{\circ}},42.0_{\ -6.6^{\circ}}^{\circ+7.2^{\circ}})$ in equatorial coordinates, which corresponds to the dipole amplitude $D=v_0/c=(1.1\pm 0.1)\times 10^{-3}$, and dipole direction $(l,b)=(180^{\circ},45^{\circ})$. The well-known CMB dipole is $v_0=369\pm 0.9~{\rm km/s}$, pointing towards $(l,b)=(264^{\circ},48^{\circ})$ \cite{Saha:2021bay}. Although the amplitudes of peculiar velocity derived from the Pantheon+ is well consistent with the CMB dipole, the direction is significantly different. They also investigated the dipole anisotropy using subsamples of Pantheon+, with different low-redshift cutoffs $z_{\rm cut}$ (note that this is different from our paper, where we using the high-redshift cutoff, rather than low-redshift cutoff.) They found that the dipole anisotropy is significant only at $z_{\rm cut}\leq 0.05$. This is consistent with our conclusion that the anisotropy only exists at low redshift region.

\vspace{5mm}
\centerline{\rule{80mm}{0.5pt}}
\vspace{2mm}

\bibliographystyle{cpc-hepnp-0}
\bibliography{reference}

\begin{thebibliography}{10}
\expandafter\ifx\csname url\endcsname\relax
  \def\url#1{\texttt{#1}}\fi
\expandafter\ifx\csname urlprefix\endcsname\relax\def\urlprefix{URL }\fi
\expandafter\ifx\csname href\endcsname\relax
  \def\href#1#2{#2} \def\path#1{#1}\fi

\bibitem{WMAP:2012fli}
C.~L. Bennett, et~al., {Nine-Year Wilkinson Microwave Anisotropy Probe (WMAP)
  Observations: Final Maps and Results}, Astrophys. J. Suppl. 208 (2013) 20.

\bibitem{WMAP:2012nax}
G.~Hinshaw, et~al., {Nine-Year Wilkinson Microwave Anisotropy Probe (WMAP)
  Observations: Cosmological Parameter Results}, Astrophys. J. Suppl. 208
  (2013) 19.

\bibitem{Planck:2013pxb}
P.~A.~R. Ade, et~al., {Planck 2013 results. XVI. Cosmological parameters},
  Astron. Astrophys. 571 (2014) A16.

\bibitem{Planck:2015fie}
P.~A.~R. Ade, et~al., {Planck 2015 results. XIII. Cosmological parameters},
  Astron. Astrophys. 594 (2016) A13.

\bibitem{Hutsemekers:2005iz}
D.~Hutsemekers, R.~Cabanac, H.~Lamy, D.~Sluse, {Mapping extreme-scale
  alignments of quasar polarization vectors}, Astron. Astrophys. 441 (2005)
  915--930.

\bibitem{King:2012id}
J.~A. King, J.~K. Webb, M.~T. Murphy, V.~V. Flambaum, R.~F. Carswell, M.~B.
  Bainbridge, M.~R. Wilczynska, F.~E. Koch, {Spatial variation in the
  fine-structure constant -- new results from VLT/UVES}, Mon. Not. Roy. Astron.
  Soc. 422 (2012) 3370--3413.

\bibitem{Mariano:2012wx}
A.~Mariano, L.~Perivolaropoulos, {Is there correlation between Fine Structure
  and Dark Energy Cosmic Dipoles?}, Phys. Rev. D 86 (2012) 083517.

\bibitem{Lineweaver:1996xa}
C.~H. Lineweaver, L.~Tenorio, G.~F. Smoot, P.~Keegstra, A.~J. Banday, P.~Lubin,
  {The dipole observed in the COBE DMR four-year data}, Astrophys. J. 470
  (1996) 38--42.

\bibitem{Tegmark:2003ve}
M.~Tegmark, A.~de~Oliveira-Costa, A.~Hamilton, {A high resolution foreground
  cleaned CMB map from WMAP}, Phys. Rev. D 68 (2003) 123523.

\bibitem{Frommert:2009qw}
M.~Frommert, T.~A. En\ss{}lin, {The axis of evil - a polarization perspective},
  PoS Cosmology2009 (2009) 015.

\bibitem{Planck:2013kqc}
N.~Aghanim, et~al., {Planck 2013 results. XXVII. Doppler boosting of the CMB:
  Eppur si muove}, Astron. Astrophys. 571 (2014) A27.

\bibitem{Saha:2021bay}
S.~Saha, S.~Shaikh, S.~Mukherjee, T.~Souradeep, B.~D. Wandelt, {Bayesian
  estimation of our local motion from the Planck-2018 CMB temperature map},
  JCAP 10 (2021) 072.

\bibitem{Riess:2018uxu}
A.~G. Riess, et~al., {New Parallaxes of Galactic Cepheids from Spatially
  Scanning the Hubble Space Telescope: Implications for the Hubble Constant},
  Astrophys. J. 855~(2) (2018) 136.

\bibitem{Riess:2021jrx}
A.~G. Riess, et~al., {A Comprehensive Measurement of the Local Value of the
  Hubble Constant with 1 km s$^{-1}$ Mpc$^{-1}$ Uncertainty from the Hubble
  Space Telescope and the SH0ES Team}, Astrophys. J. Lett. 934~(1) (2022) L7.

\bibitem{Bonvin:2006en}
C.~Bonvin, R.~Durrer, M.~Kunz, {The dipole of the luminosity distance: a direct
  measure of H(z)}, Phys. Rev. Lett. 96 (2006) 191302.

\bibitem{Antoniou:2010gw}
I.~Antoniou, L.~Perivolaropoulos, {Searching for a Cosmological Preferred Axis:
  Union2 Data Analysis and Comparison with Other Probes}, JCAP 12 (2010) 012.

\bibitem{Cai:2013lja}
R.-G. Cai, Y.-Z. Ma, B.~Tang, Z.-L. Tuo, {Constraining the anisotropic
  expansion of the Universe}, Phys. Rev. D 87~(12) (2013) 123522.

\bibitem{Zhao:2013yaa}
W.~Zhao, P.~X. Wu, Y.~Zhang, {Anisotropy of Cosmic Acceleration}, Int. J. Mod.
  Phys. D 22 (2013) 1350060.

\bibitem{Chang:2014nca}
Z.~Chang, H.-N. Lin, {Comparison between hemisphere comparison method and
  dipole-fitting method in tracing the anisotropic expansion of the Universe
  use the Union2 dataset}, Mon. Not. Roy. Astron. Soc. 446 (2015) 2952--2958.

\bibitem{Chang:2014wpa}
Z.~Chang, X.~Li, H.-N. Lin, S.~Wang, {Constraining anisotropy of the universe
  from different groups of type-Ia supernovae}, Eur. Phys. J. C 74 (2014) 2821.

\bibitem{Wang:2014vqa}
J.~S. Wang, F.~Y. Wang, {Probing the anisotropic expansion from supernovae and
  GRBs in a model-independent way}, Mon. Not. Roy. Astron. Soc. 443~(2) (2014)
  1680--1687.

\bibitem{Lin:2016jqp}
H.-N. Lin, X.~Li, Z.~Chang, {The significance of anisotropic signals hiding in
  the type Ia supernovae}, Mon. Not. Roy. Astron. Soc. 460~(1) (2016) 617--626.

\bibitem{Lin:2015rza}
H.-N. Lin, S.~Wang, Z.~Chang, X.~Li, {Testing the isotropy of the Universe by
  using the JLA compilation of type-Ia supernovae}, Mon. Not. Roy. Astron. Soc.
  456~(2) (2016) 1881--1885.

\bibitem{Wang:2017ezt}
Y.~Y. Wang, F.~Y. Wang, {Testing the isotropy of the Universe with type Ia
  supernovae in a model-independent way}, Mon. Not. Roy. Astron. Soc. 474~(3)
  (2018) 3516--3522.

\bibitem{Andrade:2017iam}
U.~Andrade, C.~A.~P. Bengaly, J.~S. Alcaniz, B.~Santos, {Isotropy of low
  redshift type Ia Supernovae: A Bayesian analysis}, Phys. Rev. D 97~(8) (2018)
  083518.

\bibitem{Chang:2017bbi}
Z.~Chang, H.-N. Lin, Y.~Sang, S.~Wang, {A tomographic test of cosmological
  principle using the JLA compilation of type Ia supernovae}, Mon. Not. Roy.
  Astron. Soc. 478~(3) (2018) 3633--3639.

\bibitem{Deng:2018yhb}
H.-K. Deng, H.~Wei, {Testing the Cosmic Anisotropy with Supernovae Data:
  Hemisphere Comparison and Dipole Fitting}, Phys. Rev. D 97~(12) (2018)
  123515.

\bibitem{Colin:2019opb}
J.~Colin, R.~Mohayaee, M.~Rameez, S.~Sarkar, {Evidence for anisotropy of cosmic
  acceleration}, Astron. Astrophys. 631 (2019) L13.

\bibitem{Chang:2019utc}
Z.~Chang, D.~Zhao, Y.~Zhou, {Constraining the anisotropy of the Universe with
  the Pantheon supernovae sample}, Chin. Phys. C 43~(12) (2019) 125102.

\bibitem{Horstmann:2021jjg}
N.~Horstmann, Y.~Pietschke, D.~J. Schwarz, {Inference of the cosmic rest-frame
  from supernovae Ia}, Astron. Astrophys. 668 (2022) A34.

\bibitem{Scolnic:2021amr}
D.~Scolnic, et~al., {The Pantheon+ Analysis: The Full Data Set and Light-curve
  Release}, Astrophys. J. 938~(2) (2022) 113.

\bibitem{Sorrenti:2022zat}
F.~Sorrenti, R.~Durrer, M.~Kunz, {The Dipole of the Pantheon+SH0ES Data},
  arXiv:2212.10328.

\bibitem{McConville:2023xav}
R.~McConville, E.~O. Colg\'ain, {Anisotropic Distance Ladder in Pantheon+
  Supernovae}, arXiv:2304.02718.

\bibitem{Pan-STARRS1:2017jku}
D.~M. Scolnic, et~al., {The Complete Light-curve Sample of Spectroscopically
  Confirmed SNe Ia from Pan-STARRS1 and Cosmological Constraints from the
  Combined Pantheon Sample}, Astrophys. J. 859~(2) (2018) 101.

\bibitem{Tripp:1997wt}
R.~Tripp, {A Two-parameter luminosity correction for type Ia supernovae},
  Astron. Astrophys. 331 (1998) 815--820.

\bibitem{Kessler:2016uwi}
R.~Kessler, D.~Scolnic, {Correcting Type Ia Supernova Distances for Selection
  Biases and Contamination in Photometrically Identified Samples}, Astrophys.
  J. 836~(1) (2017) 56.

\bibitem{duffett-smith_1989}
P.~Duffett-Smith, Practical Astronomy with your Calculator, 3rd Edition,
  Cambridge University Press, 1989.

\bibitem{Brout2022}
D.~Brout, D.~Scolnic, B.~Popovic, et~al., {The Pantheon+ Analysis: Cosmological
  Constraints}, The American Astronomical Society 938~(2) (2022) 110.

\bibitem{Foreman-Mackey:2012any}
D.~Foreman-Mackey, D.~W. Hogg, D.~Lang, J.~Goodman, {emcee: The MCMC Hammer},
  Publ. Astron. Soc. Pac. 125 (2013) 306--312.

\bibitem{Zhao:2019azy}
D.~Zhao, Y.~Zhou, Z.~Chang, {Anisotropy of the Universe via the Pantheon
  supernovae sample revisited}, Mon. Not. Roy. Astron. Soc. 486~(4) (2019)
  5679--5689.

\end{thebibliography}

\end{document}